\documentclass[letterpaper,10pt,twocolumn]{article}

\usepackage{graphicx}
\usepackage{amsmath}
\usepackage{amssymb}
\usepackage{psfrag}

\newcommand{\dd}{\textrm{d}}

\title{Area spectrum of the $D$-dimensional de Sitter spacetime}

\author{A.~L\'opez-Ortega\thanks{alopezo ``at'' ipn.mx} \\
		Centro de Investigaci\'on en Ciencia Aplicada y Tecnolog\'{\i}a Avanzada. \\
	      Unidad Legaria. Instituto Polit\'ecnico Nacional. \\
             Calzada Legaria \# 694. Colonia Irrigaci\'on. Delegaci\'on Miguel Hidalgo. \\
	      M\'exico, D.\ F., M\'exico. \\
	      C.\ P.\  11500\\ 
               }

\begin{document}

\maketitle

\begin{abstract}

The determination of the quantum area spectrum of a black hole horizon by means of its asymptotic quasinormal frequencies has been explored recently. We believe that for $D$-dimensional de Sitter horizon we must study if the idea works. Thus taking into account the local description of the thermodynamics of horizons proposed by Padmanabhan and the results of Hod, Kunstatter, and Maggiore we study the area spectrum of the $D$-dimensional de Sitter horizon.

\end{abstract}



\section{Introduction}
\label{section 1}

Supposing that the horizon area of a black hole is an adiabatic invariant and taking into account Ehrenfest principle, in 1974 Bekenstein proposed that the horizon area of a black hole in equilibrium has a discrete and equally spaced spectrum of the form \cite{Bekenstein:1974jk}--\cite{Bekenstein:1998aw}
\begin{equation}
 A_n = \epsilon \hbar n, \qquad \quad n=0,1,2,\dots,
\end{equation} 
where $\epsilon$ is a coefficient of order $1$. There are several proposals for the value of $\epsilon$, among these $\epsilon = 8 \pi$ and $\epsilon = 4 \ln (k)$, where $k$ is an positive integer, frequently appear in the literature. See \cite{Bekenstein:1995ju}--\cite{Kothawala:2008in} for some references.

The exact value of the quantity $\epsilon$ must be determined by a quantum theory of the gravity. Nevertheless the computation of $\epsilon$ by means of semiclassical methods has been previously explored. In this research line, in Ref.\ \cite{Hod:1998vk} Hod suggested that in the semiclassical limit the quantum of the black hole area can be determined from the asymptotic value of the real part of the complex quasinormal frequencies (QNF).\footnote{The QNF are the oscillation frequencies of a field that satisfies the radiation boundary conditions of a spacetime. For extensive reviews on the computation and application of the QNF in several research lines see Refs.\ \cite{Kokkotas:1999bd}--\cite{Nollert:1999bd}. } This proposal is usually known as Hod's conjecture. 

Using the spectrum of the quasinormal modes (QNM) \cite{Kokkotas:1999bd}--\cite{Nollert:1999bd} and Bohr's correspondence principle Hod was able to deduce the value of $4 \ln(3) \hbar $ for the quantum of area for four-dimensional Schwarzschild black hole \cite{Hod:1998vk}. This value is according to a rigorous statistical interpretation of its entropy \cite{Bekenstein:1997bt}--\cite{Mukhanov:1986me}, \cite{Hod:1998vk}, \cite{Kunstatter:2002pj}. Owing to these facts, Hod's proposal increases the interest in the search for a statistical derivation of the black hole entropy and in the quantization of the corresponding horizon area. See for example Refs.\ \cite{Setare:2004uu}--\cite{Polychronakos:2003js} where these ideas are explored.

Based on Hod's conjecture, in Ref.\ \cite{Kunstatter:2002pj} Kunstatter explains a method that allows to fix the spacing of the area spectrum for Schwarzschild black hole from the asymptotic value of its QNF. Key points in Kunstatter's analysis are Hod's conjecture, the first law of thermodynamics for the spacetime under study, and Bohr-Sommerfeld quantization of a classical adiabatic invariant \cite{Kunstatter:2002pj}.

Nevertheless Hod's conjecture has found some difficulties \cite{Maggiore:2007nq}, \cite{Khriplovich:2005wf}. For example, in Ref.\ \cite{Hod:1998vk} Hod considered transitions from the ground state to excited states with large $n$ \cite{Maggiore:2007nq}. Also the asymptotic value of the real part of the QNF is not universal, it can depend on the field type and spacetime studied \cite{Maggiore:2007nq}, \cite{Khriplovich:2005wf}. Another problem is that for Kerr black hole it predicts a discrete but not equally spaced area spectrum \cite{Setare:2004uu}, however for this black hole we expect an equally spaced area spectrum \cite{Bekenstein:1974jk}--\cite{Bekenstein:1998aw}, \cite{Gour:2002ga}. 

To overcome some of these difficulties, in Ref.\ \cite{Maggiore:2007nq} Maggiore suggested that the QNM of a black hole can be described as a set of damped harmonic oscillators. Based on this suggestion and on Hod's ideas, Maggiore proposed that in the semiclassical limit the quantum of area for black holes is determined by the sometimes called the physical frequency (see below) \cite{Maggiore:2007nq}
\begin{equation} \label{eq: Maggiore frequency}
 \omega_{0,n} = \sqrt{\omega_{R,n}^2 + \omega_{I,n}^2},
\end{equation} 
where $\omega_{R,n}$ and $\omega_{I,n}$ stand for the real and imaginary parts of the QNF for the black hole. Using this proposal Maggiore found that the area quantum of the Schwarzschild black hole is equal to $8 \pi \hbar$ (that is $\epsilon = 8 \pi$, see below). This value coincides with that previously obtained by Bekenstein and others using different methods \cite{Bekenstein:1974jk}--\cite{Bekenstein:1998aw}, \cite{Barvinsky:1996hr}--\cite{Padmanabhan:2003ub}, \cite{Maggiore:2007nq}, \cite{Wei:2009yj}--\cite{Medved:2008iq}. 

Recently the consequences of Maggiore's proposal have been studied in the following spacetimes, a) four-dimensional Schwarzschild black hole \cite{Maggiore:2007nq}, \cite{Wei:2009yj}, b) four-dimensional Kerr spacetime \cite{Vagenas:2008yi}, \cite{Medved:2008iq}, c) three-dimensional static BTZ black hole \cite{Wei:2009yj}, d) five-dimensional static Gauss-Bonnet black hole \cite{Wei:2009yj} (see also  \cite{Kothawala:2008in}),  e) four-di\-men\-sion\-al near extreme Schwarzschild de Sitter black hole \cite{Li:2009}, f) three-dimensional rotating black hole of the Ein\-stein-Maxwell-dilaton with cosmological constant theory \cite{Fernando:2009tv}, g) three-dimensional and five-dimensional large anti de Sitter black holes \cite{Wei:2009sw}.

Some of the results obtained in Refs.\ \cite{Maggiore:2007nq},  \cite{Wei:2009yj}--\cite{Medved:2008iq} are the following, a) the quantum of area is equal for four-dimensional Schwarzschild and Kerr black holes \cite{Maggiore:2007nq}, \cite{Wei:2009yj}--\cite{Medved:2008iq}, b) for three-dimensional static BTZ black hole the quantum of area depends on the cosmological constant \cite{Wei:2009yj}, c) the quantum of area for black hole horizons can be different for gravity theories distinct from general relativity \cite{Wei:2009yj}.

The de Sitter spacetime is a maximally symmetric solution to Einstein's equations. This simple solution has been studied thoroughly (see for example the reviews \cite{Spradlin:2001pw}, \cite{Kim:2002uz}), and it is known that the de Sitter spacetime has a horizon. According to Gibbons and Hawking the de Sitter horizon, like to the black hole horizon, is endowed with a temperature and entropy \cite{Gibbons:1977mu}. Nevertheless the de Sitter horizon, unlike to the black hole horizon, is observer dependent and the observer is located ``inside the horizon'' \cite{Spradlin:2001pw}, \cite{Kim:2002uz}. 

Owing to the observer dependence of the de Sitter horizon, it is usually believed that the de Sitter spacetime does not satisfies the analogous to the laws of black hole thermodynamics. Nevertheless, according to Padmanabhan some thermodynamical properties of the de Sitter and black hole horizons can be studied on the same basis, for example, their temperatures, their entropies, and the first law of thermodynamics \cite{Padmanabhan:2002sha}--\cite{Padmanabhan:2002ha}. For our aims we only comment that the local description of the thermodynamics of horizons given in Refs.\ \cite{Padmanabhan:2002sha}--\cite{Padmanabhan:2002ha} provides a consistent interpretation of the relation $T \dd S = \dd E + P \dd V $ in de Sitter spacetime, and we use this formulation in Sections \ref{section 2} and \ref{section 3}.

In this paper, taking into account the suggestion that the asymptotic QNF determine the area spectrum of a horizon we calculate the area quantum corresponding to the de Sitter horizon. In order to compute this quantum of area we use Padmanabhan's results on the first law of thermodynamics in de Sitter spacetime \cite{Padmanabhan:2002sha}--\cite{Padmanabhan:2002ha}, Hod's ideas \cite{Hod:1998vk}, Kunstatter's method \cite{Kunstatter:2002pj}, and Maggiore's proposal \cite{Maggiore:2007nq}. 

Among the reasons we have to study this question we enumerate the following, a) to investigate if the ideas by Hod, Kunstatter, and Maggiore work for de Sitter horizon, which in some details is different from the black hole horizon, b) if these suggestions work, then to find the quantum of area for de Sitter horizon and compare it with the quantum of area corresponding to black hole horizons, c) to study if the quantum of area for de Sitter spacetime depends on the value of the cosmological constant, d) to analyze if the quantum of area for de Sitter spacetime depends on the spacetime dimension.

The remainder of the paper is organized as follows. Following Refs.\ \cite{Padmanabhan:2002sha}--\cite{Padmanabhan:2002ha} in Section \ref{section 2} we expound a consistent form of the first law of thermodynamics for de Sitter spacetime. In Section \ref{section 3} we use the ideas by Hod, Kunstatter, and Maggiore to calculate the quantum of area corresponding to the de Sitter horizon. Finally in Section \ref{section 4} we discuss the obtained results.

\section{First law of thermodynamics for de Sitter spacetime}
\label{section 2}

In this paper we calculate the area spectrum of the $D$-dimensional de Sitter horizon, hence here we enumerate some physical properties of this spacetime. In static coordinates the $D$-dimensional de Sitter metric takes the form\footnote{Notice that the static coordinates cover the region of the de Sitter spacetime that an observer can probe. } \cite{Spradlin:2001pw}, \cite{Kim:2002uz}
\begin{equation} \label{eq: dS metric}
 \dd s^2 = - \left(1 - \frac{r^2}{L^2} \right) \dd t^2 + \left(1 - \frac{r^2}{L^2} \right)^{-1} \dd r^2 + r^2 \dd \Omega^2_{D-2},
\end{equation} 
where $ \dd \Omega^2_{D-2}$ is the line element of the unit $(D-2)$-dimensional sphere, $L$ is known as the curvature radius and is related to the cosmological constant $\Lambda$ by the expression
\begin{equation}
 L^2 = \frac{(D-1)(D-2)}{2 \Lambda}.
\end{equation} 
As is well known in de Sitter spacetime (\ref{eq: dS metric}), an observer at $r=0$ is sourrounded by a horizon located at $r=L$ \cite{Spradlin:2001pw}, \cite{Kim:2002uz}.

For de Sitter horizon we use the formulation of the first law of thermodynamics expounded in Refs.\  \cite{Padmanabhan:2002sha}--\cite{Padmanabhan:2002ha}, because it provides consistent definitions of temperature, entropy, and energy for de Sitter spacetime. According to Padmanabhan  the first law of thermodynamics for de Sitter horizon takes the form  \cite{Padmanabhan:2002sha}--\cite{Padmanabhan:2002ha}
\begin{equation} \label{eq: first law dS}
 T \dd S = \dd E + P \dd V ,
\end{equation} 
where $T$ is the Hawking temperature \cite{Gibbons:1977mu}
\begin{equation}
 T = \frac{\hbar}{2 \pi L},
\end{equation} 
$S$ stands for de Sitter entropy, that according to Bekenstein-Hawking area-entropy law is related to the area $A$ by \cite{Spradlin:2001pw}, \cite{Kim:2002uz}
\begin{equation} \label{eq: entropy dS}
 A  = \frac{2 \pi^{(D-1)/2} }{\Gamma\left( \frac{D-1}{2} \right)} L^{D-2} = 4 \hbar S ,
\end{equation} 
$V$ is the volume of the accessible region to an observer in de Sitter spacetime
\begin{equation} \label{eq: volume dS}
 V = \frac{\pi^{(D-1)/2}}{\Gamma\left( \frac{D+1}{2} \right)} L^{D-1},
\end{equation}  
$P$ is the constant pressure, and $E$ is the energy
\begin{equation} \label{eq: energy dS}
 E = - \frac{\pi^{(D-3)/2} (D-2)}{8 \Gamma\left( \frac{D-1}{2} \right) } L^{D-3}.
\end{equation} 
We notice that $\Gamma(x)$ stands for the gamma function.

The term $P \dd V$ in the thermodynamical relation (\ref{eq: first law dS}) is added for consistency because the de Sitter spacetime is a non-vacuum solution to the equations of general relativity with a source having nonzero pressure \cite{Padmanabhan:2002sha}--\cite{Padmanabhan:2002ha}.

Notice that in Refs.\ \cite{Padmanabhan:2002sha}--\cite{Padmanabhan:2002ha} is studied the four-dimensional de Sitter spacetime. Here following Section 2 of Ref.\ \cite{Paranjape:2006ca} we write the results for $D$-dimensional de Sitter spacetime. For $D=4$ the formulas (\ref{eq: entropy dS})--(\ref{eq: energy dS}) reduce to those of the references \cite{Padmanabhan:2002sha}--\cite{Padmanabhan:2002ha}.

\section{Quantum of area for $D$-dimensional de Sitter horizon}
\label{section 3}

In this section our aim is to use the ideas by Hod, Kunstatter, and Maggiore to find the quantum of area for $D$-dimensional de Sitter horizon, hence we note that the QNF of the gravitational perturbations (GP) in $D$-dimensional de Sitter spacetime are equal to \cite{Natario:2004jd}, \cite{LopezOrtega:2006my}
\begin{align} \label{eq: QNF dS gravitational}
 \omega & = - \frac{i}{L}(l + D -1 - q + 2 n),  \nonumber \\
\omega & = - \frac{i}{L}(l + q + 2n),
\end{align} 
where $n = 0, 1,\dots $, $q=0$ for tensor type GP, $q=1$ for vector type GP, $q=2$ for scalar type GP, and $l$ is the orbital angular momentum number.\footnote{The QNF of the boson and fermion fields in $D$-dimensional de Sitter spacetime for $D \geq 3$ were calculated exactly in Refs.\ \cite{Natario:2004jd}--\cite{LopezOrtega:2006ig}. It is convenient to notice that in Ref.\ \cite{Natario:2004jd} it was stated that the QNF of the de Sitter spacetime are well defined only in odd dimensions (see also Appendix A of Ref.\ \cite{Choudhury:2003wd}). In Refs.\ \cite{LopezOrtega:2006my}, \cite{LopezOrtega:2007sr} it was shown that the QNF of the de Sitter spacetime are well defined in odd and even dimensions. Also we point out that the formulas (\ref{eq: QNF dS gravitational}) determine the QNF for the massless Klein-Gordon field ($q=0$), the mode II of the electromagnetic field ($q=1$), and the mode I of the electromagnetic field ($q=2$) \cite{LopezOrtega:2006my}.} From the previous formulas for the QNF of the de Sitter background we find that $\omega_{R,n} = 0$, hence we believe that Hod's conjecture does not work in this spacetime. 

It is convenient to note that for odd dimensional de Sitter spacetimes the first formula in Eqs.\ (\ref{eq: QNF dS gravitational}) gives the QNF that already produced the second formula, thus these formulas are redundant in odd dimensions. For even dimensional de Sitter spacetimes the two formulas in  Eqs.\ (\ref{eq: QNF dS gravitational}) produce different QNF, thus in even dimensions we have two sets of QNF.

For even and odd dimensions, in the asymptotic limit $n\to \infty$, the QNF of the de Sitter spacetime (\ref{eq: QNF dS gravitational}) have identical behavior and reduce to
\begin{equation}
 \omega \approx - i\frac{2 n}{L} .
\end{equation} 
Thus we find that for de Sitter spacetime the physical frequency $\omega_{0,n}$ of the formula (\ref{eq: Maggiore frequency}), in the asymptotic limit, is given by
\begin{equation} \label{eq: physical frequency Sitter}
 \omega_{0,n} = \frac{2n}{L}.
\end{equation} 

First, taking into account that the area of the de Sitter horizon is an adiabatic invariant \cite{Mayo:1998ah}, Hod's ideas \cite{Hod:1998vk}, and Maggiore's proposal \cite{Maggiore:2007nq} here we calculate the quantum of the de Sitter horizon area spectrum. According to Maggiore, under a perturbation, the change in the mass of a black hole $\Delta M$ is given by $\Delta M = \hbar \Delta \omega$, where $\Delta \omega$ is identified with the difference between adjacent physical frequencies $\omega_{0,n}$ of the formula (\ref{eq: Maggiore frequency}) in the asymptotic limit $n \to \infty$. 

Hence we propose that a change in the parameters of the de Sitter spacetime is determined by
\begin{equation} \label{eq: proposal dS}
 \dd E + P \dd V = \hbar \Delta \omega ,
\end{equation} 
where we include the additional term $P \dd V$ in left hand side of the formula (\ref{eq: proposal dS}) because we expect that a variation in the energy of the de Sitter spacetime also include a work term. Similar to the black holes, in de Sitter spacetime we propose that $\Delta \omega$ is the difference between adjacent physical frequencies in the asymptotic limit. Therefore from the formula (\ref{eq: physical frequency Sitter}) we get
\begin{equation} \label{eq: delta omega dS}
 \Delta \omega =\omega_{0,n+1} - \omega_{0,n} = \frac{2}{L} = \frac{4 \pi T}{\hbar} .
\end{equation} 

From the thermodynamical relation (\ref{eq: first law dS}), and the formulas (\ref{eq: entropy dS}), (\ref{eq: proposal dS}), and (\ref{eq: delta omega dS}) we get
\begin{equation}
 \hbar \frac{2}{L} = T \frac{\dd A}{4 \hbar} = \frac{1}{2 \pi L} \frac{\dd A}{4 }.
\end{equation} 
Therefore, following Hod and Maggiore, we find that the quantum of area for $D$-dimensional de Sitter horizon is equal to
\begin{equation} \label{eq: quantum area dS}
 \Delta A = 16 \pi \hbar ,
\end{equation} 
thus $\epsilon = 16 \pi$ for $D$-dimensional de Sitter horizon.

Now with Kunstatter's method and Maggiore's proposal we calculate the quantum of area for de Sitter horizon, but before we expound the case of the four-dimensional Schwarzschild black hole to show how it works \cite{Maggiore:2007nq}, \cite{Wei:2009yj}.  It is well known that for a system of energy $E$ and characteristic frequency $\Delta \omega$ the quantity
\begin{equation} \label{eq: adiabatic invariant}
 I = \int \frac{\dd E}{\Delta \omega} ,
\end{equation} 
is an adiabatic invariant \cite{Kunstatter:2002pj}. Also, in the semiclassical limit, Bohr-Sommerfeld quantization states that an adiabatic invariant has an equally spaced spectrum
\begin{equation} \label{eq: I spectrum}
 I = n \hbar .
\end{equation} 

From the expression for the asymptotic QNF of the Schwarzschild black hole \cite{Motl:2002hd}--\cite{Andersson:2003fh}, for the difference between adjacent physical frequencies in the asymptotic limit we get \cite{Maggiore:2007nq}--\cite{Wei:2009yj}
\begin{equation}
 \Delta \omega = \frac{1}{4M}.
\end{equation} 
In the semiclassical limit the adiabatic invariant $I$ satisfies the formula (\ref{eq: I spectrum}), hence we get
\begin{equation}
 I = \frac{A}{8 \pi} = n \hbar,
\end{equation} 
and finally that the quantum of area for Schwarzschild black hole is equal to \cite{Maggiore:2007nq}, \cite{Wei:2009yj}
\begin{equation} \label{eq: quantum Schwarzschild}
 \Delta A = 8 \pi \hbar ,
\end{equation} 
as we previously mentioned.

Using Maggiore's proposal \cite{Maggiore:2007nq} and the ideas by Kunstatter \cite{Kunstatter:2002pj} we calculate the quantum of area for de Sitter horizon as follows. From the results of Padmanabhan on the thermodynamics of the $D$-dimensional de Sitter horizon previously expounded in Section \ref{section 2} and following Setare and Vagenas \cite{Setare:2004uu}, Vagenas \cite{Vagenas:2008yi}, and Medved \cite{Medved:2008iq}, for de Sitter spacetime the analogous to the adiabatic invariant $I$ of the formula (\ref{eq: adiabatic invariant}) is 
\begin{equation} \label{eq: adiabatic dS}
 I = \int \frac{\dd E + P \dd V}{\Delta \omega},
\end{equation} 
where we take $\Delta \omega$ as in the formula (\ref{eq: delta omega dS}).

From the formulas of $E$,  $P$, and $V$ for $D$-dimensional de Sitter spacetime given in Section \ref{section 2} we obtain that the postulated adiabatic invariant (\ref{eq: adiabatic dS}) is equal to  
\begin{align}
 I & = \frac{\pi^{(D-3)/2} (D-2)}{4 \Gamma \left(\frac{D-1}{2} \right) } \frac{1}{2} \int L^{D-3} \dd L \nonumber \\
& = \frac{1}{16 \pi} \frac{2 \pi^{(D-1)/2} L^{D-2}}{ \Gamma\left( \frac{D-1}{2} \right)}.
\end{align}
Taking into account the expression for the area of the $D$-dimensional de Sitter horizon (\ref{eq: entropy dS}) we find 
\begin{equation}
 I = \frac{A}{16 \pi},
\end{equation} 
and therefore, from the formula (\ref{eq: I spectrum}), we get that the area spectrum of the de Sitter horizon is equal to 
\begin{equation}  \label{eq: discrete area dS}
 A_n = 16 \pi \hbar n .
\end{equation} 

Thus, from this formula, we find that the spacing of the area spectrum for $D$-dimensional de Sitter spacetime takes the same form given in the formula (\ref{eq: quantum area dS}), already calculated with Hod's proposal and Maggiore's suggestion. Similar to the Schwarzschild and Kerr black holes \cite{Maggiore:2007nq}, \cite{Wei:2009yj}--\cite{Medved:2008iq}, both methods give identical results for the quantum of area for $D$-dimensional de Sitter spacetime. Owing to Bekenstein-Hawking area-entropy law (\ref{eq: entropy dS}), the spacing of the entropy spectrum for $D$-dimensional de Sitter horizon is 
\begin{equation} \label{eq: quantum entropy dS}
 \Delta S = 4 \pi . 
\end{equation} 

Notice that the expression (\ref{eq: quantum area dS}) for the quantum of area for $D$-dimensional de Sitter horizon does not depend on the spacetime dimension. In Ref.\ \cite{Wei:2009yj} it was found that for three-dimensional static BTZ black hole the quantum of area depends on the cosmological constant.\footnote{In Ref.\ \cite{Wei:2009yj} for three-dimensional static BTZ black hole was calculated an area spectrum of the form $A_n = 2 \pi \hbar n / \sqrt{\Lambda} $, hence the quantum of area is $\Delta A = 2 \pi \hbar / \sqrt{\Lambda}$, where $\Lambda$ denotes the three-dimensional cosmological constant.} For $D$-dimensional de Sitter horizon the quantum of area (\ref{eq: quantum area dS}) is independent of the cosmological constant.

As we commented before, for Schwarzschild black hole was found that the quantum of area takes the form (\ref{eq: quantum Schwarzschild}). Comparing this result with the expression (\ref{eq: quantum area dS}) we get that the area quantum for $D$-dimensional de Sitter horizon is twice as big as the corresponding to the horizon of the four-dimensional Schwarzschild black hole.

\section{Discussion}
\label{section 4}

We find that Hod's ideas, Maggiore's suggestion, and Kunstatter's method work in de Sitter spacetime and allow us to calculate the area spectrum of the de Sitter horizon. The computed value of the area quantum of the de Sitter spacetime (\ref{eq: quantum area dS}) is twice as big as the corresponding to the Schwarzschild black hole (\ref{eq: quantum Schwarzschild}). Also its value does not depend on the spacetime dimension and on the cosmological constant. Our result for the quantum of area for de Sitter spacetime points out that in general relativity the quantum of area is different for de Sitter and black hole horizons. 

For calculating the physical frequencies $\omega_{0,n}$ of the de Sitter spacetime instead of taking the asymptotic limit of the QNF as in Section \ref{section 3}, we can use the exact results of the QNF (\ref{eq: QNF dS gravitational}), with $l$ fixed. From these exact values we get that for all $n$ the difference between adjacent physical frequencies is also given by the formula (\ref{eq: delta omega dS}). Thus we get the same value of $\Delta \omega$ previously calculated in the asymptotic limit. In Section \ref{section 3} we take the asymptotic limit in order to use Bohr's correspondence principle \cite{Hod:1998vk}, however if the previous fact has other implications deserves detailed study.

In Ref.\ \cite{Li:2009} for near extremal Schwarzschild de Sitter black hole the value of its area quantum was calculated with the second method used in the previous section (see also \cite{Setare:2004rt}). The computed value in Ref.\ \cite{Li:2009} for the quantum of area is $\Delta A = 24 \pi \hbar$ (which is the sum of the area quantums corresponding to Schwarzschild and de Sitter spacetimes). However it has been shown that the near extremal charged black holes and the near extremal rotating black holes are highly quantum objects \cite{Gour:2002ga}--\cite{Gour:2002uk}. If the near extremal Schwarzschild de Sitter spacetime behaves in a similar way to the near extremal charged or rotating black holes deserves detailed study to support the results of Ref.\ \cite{Li:2009}.

\section{Acknowledgements}

I thank Dr.\ C.\ E.\ Mora Ley and A.\ Tellez Felipe for their interest in this paper. This work was supported by CONACYT M\'exico, SNI M\'exico, EDI-IPN, COFAA-IPN, and Research Project SIP-20090952.

\end{document}